\documentclass[prl,twocolumn,showpacs,superscriptaddress]{revtex4}
\usepackage{graphicx}
\usepackage{dcolumn}
\usepackage{bm}

\def\beq{\begin{equation}}
\def\bea{\begin{eqnarray}}
\def\eeq{\end{equation}}
\def\eea{\end{eqnarray}}     
\begin{document}
\title{Continuous Melting of a ``Partially Pinned'' Two-Dimensional
Vortex Lattice in a Square Array of Pinning Centers}

\author{Toby Joseph}
\email{toby@physics.iisc.ernet.in}

\author{Chandan Dasgupta}
\email{cdgupta@physics.iisc.ernet.in}
\affiliation{Centre for Condensed Matter Theory, Department of Physics,
Indian Institute of Science, Bangalore 560 012, India}

%\date{\today}

\begin{abstract}

The structure and equilibrium properties of a two-dimensional system of
superconducting vortices in a periodic pinning potential with square
symmetry are studied numerically. For a range of the strength of the
pinning potential, the low-temperature crystalline state exhibits only
one of the two basic periodicities (in the $x$- and $y$-directions) 
of the pinning potential. This ``partially pinned'' solid undergoes a 
continuous melting transition to a weakly modulated liquid as the 
temperature is increased. A spin model, constructed using symmetry 
arguments, is shown to reproduce the critical behavior at this transition.

\end{abstract}

% PACS, the Physics and Astronomy Classification Scheme.
\maketitle

Thin-film superconductors with artificially constructed periodic arrays
of pinning centers have attracted much
experimental~\cite{baert,harada,martin1} and
theoretical~\cite{reich2} attention in recent years. Such arrays
may consist of micro-holes (``antidots''~\cite{baert}), defects
produced by heavy-ion bombardment~\cite{harada}, or magnetic
dots~\cite{martin1}. The presence of a periodic pinning
potential has many interesting effects on the equilibrium and transport
properties of a system of vortices induced in the sample by an external
magnetic field. Some of these effects, which depend crucially on the value
of the filling factor $n$ that measures the average number of
vortices per unit cell of the pin lattice, have been observed in
imaging experiments~\cite{harada} and in measurements of the
magnetic~\cite{baert} and transport~\cite{martin1}
properties of such samples. The melting transition of the vortex
lattice in such systems provides an example of two-dimensional (2d)
melting in an external periodic potential. Evidence for this melting
transition has been found in imaging experiments~\cite{harada} and
magnetization measurements~\cite{baert}. Similar melting transitions
are of interest in other physical systems such as atoms adsorbed on
crystalline substrates\cite{nh}, arrays of Josephson
junctions~\cite{franz}, and colloidal particles in interfering laser
fields~\cite{laser}.
%and arrays of optical traps~\cite{traps}.

The effects of a weak, commensurate, periodic potential on 2d melting
have been studied~\cite{nh} within the framework of the
Kosterlitz-Thouless-Halperin-Nelson-Young theory~\cite{nh,kt} of defect
mediated melting. For $n\ll 1$, this analysis predicts the occurrence
of two continuous transitions: a {\it depinning} transition from a
low-temperature {\it pinned solid} phase in registry with the substrate
to a {\it floating solid} phase that is essentially decoupled from 
the substrate, and a {\it melting} transition at a higher temperature
where the floating solid transforms to a liquid. This sequence of
transitions has been observed in simulations~\cite{reich2,franz}
of 2d systems in weak, commensurate, periodic potentials with
triangular and square symmetry. 
%However, some of the simulation
%results~\cite{franz} about the nature of the transitions differ
%from the analytic predictions. 
The two transitions are expected to
merge into a single one (from the pinned solid to the liquid) as $n$ is
increased~\cite{nh,franz} and/or the pinning potential is
strengthened~\cite{reich2}. 

In this paper, we report the results of Monte Carlo (MC) simulations of
the equilibrium properties of a 2d system of vortices in the presence
of a square array of pinning centers. The filling factor $n$ is taken
to be unity. We consider pinning centers that produce a
repulsive potential with range comparable to that of the intervortex
interaction. Pinning centers with these properties may be
experimentally realized in
arrays of magnetic dots each of which produces a potential that can be
tailored~\cite{mdots} by adjusting its magnetic moment. Another physical
realization is obtained in a square array of strong, attractive,
short-range pinning centers at filling $n=2$~\cite{reich2}. In this case,
each pinning center would trap a vortex at low temperatures, and these
pinned vortices would interact with the remaining interstitial vortices
(assuming each pin can trap only one vortex) via an effective repulsive
potential~\cite{ks}. The net potential produced by an array of such
pinning centers has very flat minima~\cite{ks} at the centers of the
square unit cells of the pin lattice (see Fig.1). This kind of pinning is
qualitatively different from that considered in previous
studies~\cite{reich2,franz} in which each pinning center was assumed to
produce an attractive potential with range much smaller than the
intervortex spacing.

We find that this difference in the nature of the pinning has strong
effects on the structure of the low-temperature solid phase and its
melting transition. For a range of values of the strength of the
pinning potential, the low-temperature state of our vortex system is a
lattice with a basis, with unit cells of size $2d \times d$, where $d$
is the spacing of the pin lattice, and two vortices in each unit cell
(see Fig.1). Three other structures, related by symmetry to the one
shown in Fig.1, are equally likely to occur at low temperatures. We
call this phase, which exhibits one of the two basic periodicities
of the square pin array, a {\it partially pinned} solid to distinguish
it from the pinned and floating solid phases mentioned above. 
%The
%occurrence of a vortex structure with one of the two basic
%periodicities of the pin lattice has been
%suggested as an explanation of recent experimental
%results~\cite{martin2} on a thin-film superconductor with a rectangular
%array of magnetic dots.

As the temperature is increased, the partially pinned solid undergoes a
continuous melting transition to a weakly modulated liquid that has the
square symmetry of the substrate. We have used finite-size scaling to
analyze the critical behavior at this transition. We have also used
symmetry considerations to construct a spin model that is expected to
exhibit a transition in the same universality class as that of the
melting transition in the vortex system. The values of critical
exponents obtained from a finite-size scaling analysis of the results
of simulations of the spin model are consistent with those
obtained for the vortex system. This transition appears to belong in an
universality class not found in previous studies of similar systems.

We model the 2d system of vortices as a collection of point
particles interacting via the repulsive potential
\begin{equation} 
U(r) = U_0 K_{0}(r/\lambda).
\label{inter}
\end{equation}
Here, $K_{0}$ is the Hankel function and $U_0 =
\Phi_{0}^{2}t/(8\pi^{2}\lambda^{2})$, where $\Phi_{0}$ is the flux 
quantum, $\lambda$ is the penetration depth, and $t$ is the film thickness.
The interaction of the vortices with
the pinning centers is assumed to be of the form $AU(r)$, where the
parameter $A$ measures the relative strength of the pinning potential. 
We use parameters appropriate for the Nb sample studied in
Ref.\cite{harada}: $\lambda$ = 0.1$\mu$m, $d = 10 \lambda$, $t = \lambda$. 
For these parameter values, $U(r=d)/k_B \simeq$ 7K. 
We measure lengths in units of
$\lambda$, energies in units of $U_0$, and the temperature in Kelvins.

In Fig.1, we have shown the variation of the net pinning potential for
$A=10^{-3}$ along a diagonal of a pin square, and compared it with the
potential due to a system of vortices located at the centers of the pin
squares. The pining potential exhibits a very flat minimum at the
center of the square.  This plot also shows that the pinning and
interaction energies are comparable for this value of $A$.  We used a
simulated annealing procedure to find the ground states of the vortex
system for various values of $A$. When $A$ is of order unity or higher,
the ground state has square symmetry, with one vortex located at the
center of each elementary square of the pin array. As $A$ is reduced
below $A_{c}\simeq 0.012$, the ground state is found to be made up of
unit cells that consist of two pin squares. The vortices in the two
squares are displaced from the centers by equal amounts in opposite
directions, as shown in the lower inset of Fig.1.  This displacement
causes a reduction of the interaction energy, which more than
compensates the increase in the pinning energy if $A$ is small. The
distance $D_{min}$ of the vortices from the center of the square
increases as $A$ is decreased, and reaches a maximum of about $0.23d$
for $A \approx 10^{-4}$.  We have also calculated the energy of a
vortex lattice of this structure for different values of the
displacement $D$ from the center and found the $D$ that minimizes the
energy for different $A$. The results, shown in Fig.1, match well with
those obtained from simulated annealing. These results clearly show a
transition from a fully pinned structure with square symmetry to a
partially pinned structure with $D_{min} \ne 0$ as $A$ is decreased
below $A_c$. The ground
states for $A<10^{-5}$ appear to have a complex, disordered structure
(we are not sure that our simulated annealing procedure located the
true ground states for such values of $A$), and the triangular
Abrikosov lattice is recovered for $A=0$.

\begin{figure} \includegraphics[scale=0.35]{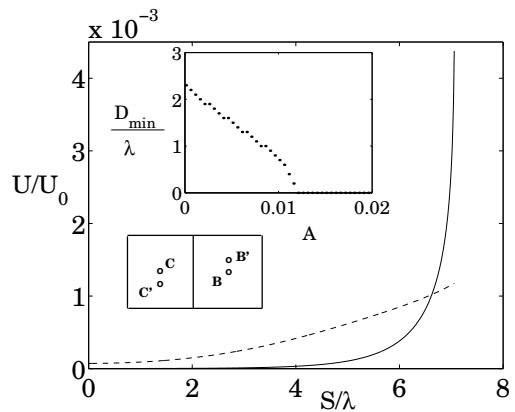}
\caption{The main plot shows the net pinning potential for $A=10^{-3}$ 
(solid line) and the potential due to vortices located at 
the centers of the squares (dashed line) as functions of the distance
$S$ from the center of a pin square along a diagonal.
%(in units of $\lambda$) 
%The potential $U$ is measured in units of $U_0$. 
The upper inset shows a plot of $D_{min}$, the displacement from the 
center that minimizes the lattice energy, as a
%(in units of $\lambda$) 
function of the pinning-strength parameter $A$. 
The lower inset shows an unit cell of
the partially pinned structure. $B$ and $C$ are the positions of the
vortices in the state with square symmetry and $B'$ and $C'$ are their
positions in the partially pinned structure, with $BB'=CC'=D_{min}$.
Three other structures, related to the shown one by symmetry, are
equally probable at low temperatures.}
\end{figure}

At high temperatures, the vortices form a weakly
modulated liquid with square symmetry. We used MC simulations
to study how the system evolves to this state as the temperature $T$ is
increased. We monitored structural changes by measuring the structure
functions $S({\bf k}) = \langle\rho({\bf k})\rho(-{\bf k})\rangle/N^2$,
where $\langle \cdots \rangle$ represents a thermodynamic (MC)
average, $\rho({\bf k})$ is the Fourier transform of the local density,
and $N$ is the number of particles in the system. We
also looked for signatures of a transition by measuring the specific
heat, $C_{v} = \langle(E-\langle E \rangle)^{2} \rangle/(Nk_{B}T^{2})$,
where $E$ is the total energy of the system. When $A$ is large
enough to have a lattice with square symmetry as the ground state, the
system gradually transforms to a modulated liquid with the same
symmetry with no signature of a phase transition. For lower values of
$A$, when the symmetry of the ground state is different from that of
the high temperature phase, the system undergoes a continuous
transition that is signalled by a peak in the specific heat and sharp
changes in $S({\bf k})$ for appropriate $\bf k$s, as shown in Fig.2
and Fig.3. 

The reciprocal lattice vectors of the structure with the unit cell
shown in Fig.1 are ${\bf G}(n_1,n_2) = (n_{1}\frac{\pi}{d},n_{2}
\frac{2\pi}{d})$, where $n_{1}$ and $n_{2}$ are integers. For this
structure, $\langle \rho({\bf G}) \rangle$ vanishes for
the smallest ${\bf G}$ corresponding to $n_1=1$, $n_2=0$. We have
measured the temperature-dependence of $S({\bf G})$ for the next three
smallest $\bf G$s: ${\bf G}_1 = {\bf G}(2,0)$, ${\bf G}_2 = {\bf
G}(1,1)$, and ${\bf G}_3 = {\bf G}(0,1)$. Note that ${\bf G}_1$ and
${\bf G}_3$ are reciprocal lattice vectors of the square pin array, but
${\bf G}_2$ is not. Simulation data obtained in a cooling run for a
sample with $A = 10^{-4}$ are shown in Fig.2. At
high temperatures, $S({\bf G}_1)$ and $S({\bf G}_3)$ have the same
small value, whereas $S({\bf G}_2)$ is smaller, indicating a weakly
modulated phase with square symmetry.
As the temperature is decreased, $S({\bf G}_1)$ and $S({\bf G}_2)$
increase sharply near $T \simeq 5.0$K, while $S({\bf G}_3)$ decreases
at about the same temperature. As the temperature is decreased further,
the values of $S({\bf G}_1)$ and $S({\bf G}_2)$ approach unity, while 
$S({\bf G}_3)$ goes to zero, indicating a partially pinned structure. 
As shown in the inset of Fig.2,
the average displacement $D$ of the vortices from the centers of the pin
squares exhibits a sharp increase as the temperature is decreased across 
$T \simeq 5.0$K. These results strongly suggest a
phase transition between the partially pinned state and a state with square 
symmetry at $T = T_c \simeq 5$K.

\begin{figure}
\includegraphics[scale=0.35]{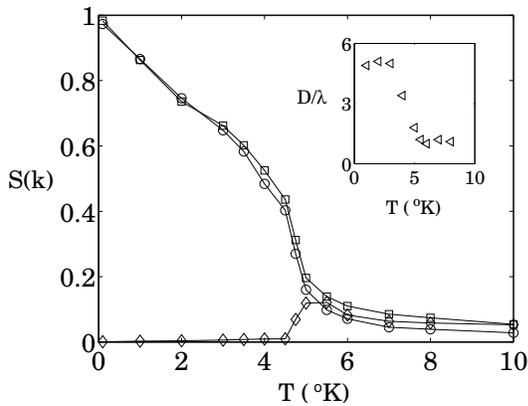}
\caption{Temperature dependence of the structure functions $S({\bf G})$ 
for $A=10^{-4}$. Data for three values of $\bf G$ (see text), ${\bf G}_1$
(squares), ${\bf G}_2$ (circles) and ${\bf G}_3$ (diamonds) are shown.
The solid lines are guides to the eye. The inset shows the average 
displacement $D$ of the vortices from the centers of the pin squares 
as a function of the temperature $T$. 
%The temperature $T$ in both
%the plots is measured in Kelvins and $D$ is measured in units of
%$\lambda$.
} 
\end{figure}

In order to determine the nature of the transition, we have carried out
a finite-size scaling study of the specific heat $C_v$ of the system.
As shown in Fig.3, $C_v$ peaks at the transition temperature obtained
from the behavior of $S({\bf G})$ and $D$, and the peak becomes higher
and sharper as the size of the system is increased. This is the
behavior expected at a continuous phase transition for which
finite-size scaling theory~\cite{fsc} predicts that the peak value of
$C_v$ should scale as $L^{\alpha/\nu}$ where $L=\sqrt{N}$ is the linear
size of the system, and $\alpha$ and $\nu$ are, respectively, the critical
exponents for the specific heat and the correlation length. As shown in
the inset of Fig.3, our data are quite consistent with this behavior,
with $\alpha/\nu \simeq 0.46$. Moreover, we have checked for hysteresis
by measuring the specific heat during heating and cooling runs. We did
not find any evidence for hysteresis, confirming that the transition is
a continuous one. The transition temperature is found~\cite{unpub} to
depend non-monotonically on the strength $A$ of the pinning potential.
%We also found~\cite{unpub} that the transition
%temperature $T_c$, operationally defined as the value of $T$ at the peak of
%$C_v$ for a fixed sample-size, does not depend monotonically on the
%strength $A$ of the pinning potential. For large values of
%$A$, only the phase with square symmetry exists and there is no
%transition. 
As $A$ is decreased below a critical value near $10^{-2}$, 
the transition temperature increases
from zero, attains a maximum near $A = 8.10^{-5}$, and then decreases as 
$A$ is decreased further.
%Further decrease in $A$
%causes $T_c$ to decrease, suggesting a {\it re-entrant} behavior as a
%a function of $A$ with $T$ fixed slightly below the maximum value of
%$T_c$. We have also carried out a density functional
%calculation~\cite{unpub} that confirms this behavior.

\begin{figure}
\includegraphics[scale=0.35]{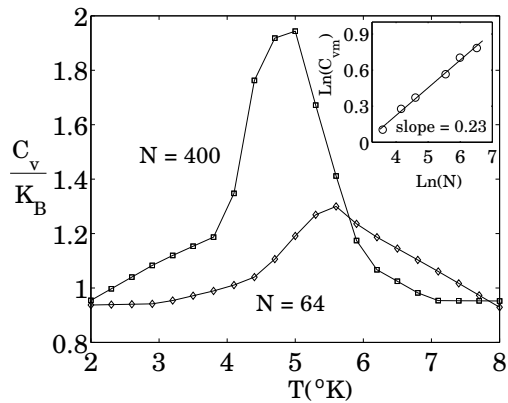}
\caption{The specific heat $C_{v}$ (in units of $k_B$) as a function of
the temperature $T$ 
%(in Kelvins)
for $A=10^{-4}$ and two sample sizes, $N=64$ and $N=400$. The lines joining
the data points are guides to the eye. The inset shows a double-log
plot of $C_{vm}$, the peak value of the specific heat, as a function of the
system size $N=L^2$.}
\end{figure} 

We have used symmetry arguments to construct a spin model that should
exhibit a phase transition in the same universality class as the
transition found in the vortex system. As noted earlier, the vortex system
has four degenerate ground states: two in which the displacements from the
centers of the pin squares are $\pm D_{min} \hat{\bf y}$ ($\mp D_{min}
\hat{\bf y}$) in even (odd) columns of the pin lattice, and two others
in which the displacements are $\pm D_{min} \hat{\bf x}$ 
($\mp D_{min} \hat{\bf x}$) in even (odd) rows ($\hat{\bf x}$ and 
$\hat{\bf y}$ are unit vectors in the horizontal and vertical
directions, respectively).
We, therefore, consider four-state planar ``spin'' variables 
${\bm \sigma}_i$ located at the sites of the dual of the pin lattice.
Each of these variables has unit length and can point in the four
directions, $\pm \hat{\bf x}$, and $\pm \hat{\bf y}$. One may think of
these variables as representing the directions of small displacements
of magnitude $\delta$ from the centers of the pin squares. The distance
$s_{ij}$ between two vortices in neighboring pin squares, with displacements
$\delta {\bm \sigma}_i$ and $\delta {\bm \sigma}_j$, is given by
\begin{equation} 
s_{ij}^{2} = d^2+2 \delta^{2}(1-{\bm \sigma}_{i}.{\bm
\sigma}_{j})+2\delta d(\sigma_{j\alpha}-\sigma_{i\alpha}),
\label{dist}
\end{equation}
where $\alpha$ is $x$ ($y$) if the spins ${\bm \sigma}_i$ and ${\bm
\sigma}_j$ are separated horizontally (vertically). Since the
intervortex interaction depends only on the distance, and the pinning
potential is {\it independent} of the orientation of the ${\bm
\sigma}$s, the symmetry of the vortex problem would be preserved in the
spin model if its
Hamiltonian is taken to be a suitably chosen function of $s^2$ that
leads to the fourfold-degenerate ground-state structures described
above. We have found that the Hamiltonian
\begin{equation}
{\mathcal H} = J_{o}\sum_{<ij>} \exp(-s_{ij}^{2}/s_0^{2}),
\label{spinmodel}
\end{equation}
where the sum is over nearest-neighbor pairs and $J_0$ is an energy
parameter, leads to the expected ground state structure if the length
parameter $s_0$ is sufficiently large. By expanding the exponential in
Eq.(\ref{spinmodel}) and making use of the properties of the
${\bm \sigma}_i$s, this Hamiltonian may be written as
\begin{eqnarray}
{\mathcal H} &=& \sum_{<ij>} [J_{1}{\sigma}_{i\alpha}{\sigma}_{j\alpha} 
 + J_{2}{\sigma}_{i\beta}{\sigma}_{j\beta} 
%\hspace*{2cm} \nonumber \\ 
 + J_{3}\sigma_{i\alpha}\sigma_{j\alpha}(\sigma_{i\alpha} 
 - \sigma_{j\alpha}) \nonumber \\
 &+& J_{4} \sigma_{i\alpha}^{2}\sigma_{j\alpha}^{2}
 + J_{5} \sigma_{i\beta}^{2}\sigma_{j\beta}^{2}],
\label{hamiltonian}
\end{eqnarray}
where $\alpha,\beta$ are $x,y$ ($y,x$) for 
horizontal (vertical) bonds, and $J_{1},J_{2},J_{3},J_{4}$ and $J_{5}$
are functions of $J_{0}, \delta, d$ and $s_0$. While the spins in our model
are analogous to those in the four-state clock model, the Hamiltonian of 
Eq.(\ref{hamiltonian}) {\it does not} have the $z(4)$ symmetry of the clock
model.

\begin{figure}
\includegraphics[scale=0.35]{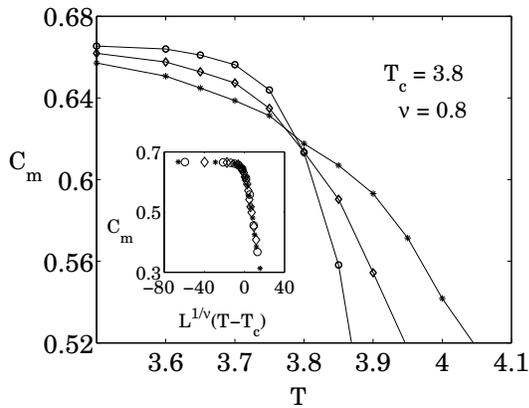}
\caption{The main plot shows the order parameter cumulant $c_m$ of the
spin model (see text) as a function
of temperature $T$ for three system sizes, $N=144$ (stars), $N=324$
(diamonds), and $N=1024$ (circles). 
%The point of intersection of the curves gives the transition
%temperature. 
The inset shows a finite-size scaling plot of $c_m$
versus $L^{1/\nu} (T-T_c)$ with $T_c=3.8$ and $\nu=0.8$.
%The temperature is in units of degree kelvin.
} 
\end{figure}

We have performed extensive MC simulations of the thermodynamics
of the spin model for $J_{0} = 10$ (this sets the temperature scale), 
$d/s_0 = 2$ and $d/\delta = 4$. For these parameter values, we find
four degenerate ground states in which the spins point in $\pm \hat{\bf
x}$ directions in alternate rows, or in $\pm \hat{\bf y}$ directions in
alternate columns. The order parameter $m$ is defined as
\begin{equation}
m = \frac{1}{N} \langle |\Sigma_1 \sigma_{ix} -\Sigma_2 \sigma_{ix}| 
+ |\Sigma_3 \sigma_{iy} -\Sigma_4 \sigma_{iy}|\rangle,
\label{op}
\end{equation}
where the four sums $\Sigma_k, k=1-4$ are over even rows, odd rows,
even columns, and odd 
columns, respectively. This definition ensures that $m=1$ in
%the absolute
%value of the difference in sub lattice magnetization composed of
%alternate rows added to a similar quantity for alternate columns.
any of the four ground states. The transition temperature $T_c$ and the
correlation-length exponent $\nu$ were determined from a finite-size
scaling analysis of the data for the Binder cumulant~\cite{fsc} for the order
parameter, $c_m \equiv 1-\langle m^4 \rangle/(3\langle m^2 \rangle^2)$. 
The $L$- and $T$-dependence of this quantity near the transition 
is expected to have the form $c_m(L,T)=f(L^{1/\nu}t)$ where $f$ is a
scaling function and $t=(T-T_c)/T_c$. As shown in Fig.4, plots of $c_m$
vs. $T$ for different sample sizes intersect at $T=T_c\simeq 3.8$,
confirming the occurrence of a continuous transition. The inset of
Fig.4 shows that the data for $c_m$ for different $L$ and $T$ collapse
to the same scaling curve when plotted against $L^{1/\nu}(T-T_c)$ with 
$\nu =0.8$. The other critical exponents, computed from finite-size
scaling analysis of specific heat, order parameter and susceptibility
data~\cite{unpub} are: $\beta \simeq 0.05$, $\gamma \simeq 1.5$ and
$\alpha \simeq 0.4$. These values are consistent, within error bars,
with the value of $\alpha/\nu$ obtained for the vortex system. These exponent
values establish that the transition in our model is not in the
universality class of the four-state clock model. The observed critical
behavior is also quite different from that expected~\cite{nh,kt} near a
Kosterlitz-Thouless-type transition. It appears that the 
universality class of this transition is different from those
found in earlier studies of similar systems.
%not aware of any 2d spin model with these exponent values.

To conclude, we have shown that a 2d system of particles in a periodic
potential with square symmetry can have a partially pinned
low-temperature phase that undergoes a continuous transition to a
weakly modulated liquid as the temperature is increased. After the submission
of this paper, we came across two recent papers that confirm some
of our predictions. An experimental study~\cite{fasano}
has analyzed the vortex structure at the bottom surface of a thin
superconductor
with a commensurate square array of pinning centers on the top surface.
Vortex lines pinned at the top surface provide, via 
their elastic energy, an effective pinning potential of square symmetry 
at the bottom surface whose strength is a decreasing function of the
sample thickness. The experiment finds a structure with the 
same symmetry as that of our ``partially pinned'' lattice for a range of
thicknesses. A numerical modeling~\cite{theory} of the experimental system
also yields results quite similar to those shown in the 
insets of Fig.1.
%We have
%constructed a four-state spin model that captures the symmetry of the
%particle system and shown that it belongs in the same universality
%class. 

%This transition should be experimentally observable in systems
%such as 2d superconductors with a square array of magnetic dots, and 2d
%colloidal systems in crossed laser fields with $n=2$, where the
%effective pinning potential for the interstitial particles can be tuned
%by varying the charge of the strongly pinned ones.

\end{document}